# Shooting the Messenger? Harassment and Hate Speech Directed at Journalists on Social Media

**Simón Peña-Fernández[1]\*, Urko Peña-Alonso[2], Ainara Larrondo-Ureta[3] and Jordi Morales-i-Gras[4]**

1. University of the Basque Country (UPV/EHU) (ROR: 000xsnr85)
   simon.pena@ehu.eus / https://orcid.org/0000-0003-2080-3241
2. University of the Basque Country (UPV/EHU) (ROR: 000xsnr85)
   urko.pena@ehu.eus / https://orcid.org/0000-0002-9214-5906
3. University of the Basque Country (UPV/EHU) (ROR: 000xsnr85)
   ainara.larrondo@ehu.eus / https://orcid.org/0000-0003-3303-4330
4. University of the Basque Country (UPV/EHU) (ROR: 000xsnr85)
   jordi.morales@ehu.eus / https://orcid.org/0000-0003-4173-3609

\* Corresponding author

**Abstract:** Journalists have incorporated social networks into their work as a standard tool, enhancing their ability to produce and disseminate information and making it easier for them to connect more directly with their audiences. However, this greater presence in the digital public sphere has also increased their exposure to harassment and hate speech, particularly in the case of women journalists. This study analyzes the presence of harassment and hate speech in responses ($n$ = 60,684) to messages that 200 journalists and media outlets posted on X (formerly Twitter) accounts during the days immediately preceding and following the July 23 (23-J) general elections held in Spain in 2023. The results indicate that the most common forms of harassment were insults and political hate, which were more frequently aimed at personal accounts than institutional ones, highlighting the significant role of political polarization—particularly during election periods—in shaping the hostility that journalists face. Moreover, although, generally speaking, the total number of harassing messages was similar for men and women, it was found that a greater number of sexist messages were aimed at women journalists, and an ideological dimension was identified in the hate speech that extremists or right-wing populists directed at them. This study corroborates that this is a minor but systemic issue, particularly from a political and gender perspective. To counteract this, the media must develop proactive policies and protective actions extending even to the individual level, where this issue usually applies.

## 1. Introduction

Social networks are an essential tool in the daily work of journalists, who use them to follow current events, contact sources, or search for new topics or focuses to report on [1,2]. They also serve as powerful instruments to broaden the reach of published content and strengthen audience engagement, always with the aim of increasing the reach and impact of their content [3,4]. The relationship between journalists and audiences is being redefined as increasingly interactive and participatory. Platforms incorporate new dynamics of social validation of journalistic knowledge [5], challenging hierarchical structures and fostering more collaborative relationships in which engagement and dialog are central to contemporary journalistic practice [6,7].

In contrast to traditional news media, where journalists adhere to established norms to uphold autonomy and authority, social media platforms operate beyond institutional boundaries. These hybrid spaces foster connection and self-presentation, blurring personal–professional and public–private divides [8]. In many newsrooms, the use of social networks has become a natural extension of journalistic work, which is carried out using not only the corporate profiles of the media outlets but also the personal accounts of the journalists themselves. In fact, four times as many professionals use both types of profiles simultaneously compared to those who rely solely on institutional accounts, revealing the extent to which social media has been integrated into journalistic practice [9].

As a result, not only can journalists achieve a much better connection with their audiences, as well as improved ways to interact more directly with them in a shared space [10,11], but it also enables the creation of their own brand image, boosting their profiles beyond the news outlets where they work [12].

However, this greater presence on social networks also implies a heavier workload, greater urgency in producing content rapidly, and heightened public exposure, which exacerbates professional risks and personal vulnerability due to the blurring of boundaries between work and private lives [13]. As Lewis and Molyneux [14] point out, the journalist's role has been reshaped not only by platform affordances but also by institutional pressures and audience expectations. Rather than empowering journalists, social media often imposes additional burdens, fosters insularity, and exposes them to harassment—challenging assumptions of enhanced engagement and calling for more critical reflection on journalistic labor and vulnerability.

In its most extreme form, the disadvantage of using social networks is the harassment of journalists, whether due to the failure of regulatory and protection standards to stop abuse, the speed with which information circulates on the Internet, the ease of maintaining anonymity, or the enormous volume of messages generated [15].
Attacks on journalists via social media have at least three important dimensions that highlight the comprehensive impact that they have on the profession. First, in terms of personal well-being, pervasive harassment can lead to anxiety and psychological problems among those experiencing it [16–20], which can lead to fatigue, burnout, and a desire to quit one's job prematurely [21].

Second, in terms of the quality of professional practice, online harassment can ultimately influence the types of stories that are posted on social networks or the ways in which potentially controversial topics are covered [22], a particularly significant repercussion in the case of a social group such as journalists, whose function is specifically to help foster debate in the public sphere [23].

Therefore, ultimately, attacks on journalists via social media represent a danger to the public as a whole in that they can threaten journalists' freedom of expression and, by extension, that of democratic societies themselves, given that they discourage or hinder participation in public debate through self-censorship or lead professionals to avoid dialogs with their audiences [24]. This type of violence also aims to undermine journalists' professional reputations and damage their credibility, thereby restricting the free practice of journalism [19] and jeopardizing public trust in the media, pluralism, and social cohesion [25].

Consequently, studies by international organizations such as Reporters Without Borders (RWB) and the United Nations Educational, Scientific and Cultural Organization (UNESCO) have reported concern regarding the harassment of journalists in the digital sphere and the ways in which it may influence their work [26,27]. The rise of populism and the stigmatization of the press as an "enemy" have further intensified these attacks [28–30]. Such rhetoric often seeks to delegitimize and silence journalists under the guise of protecting freedom of expression [31]. Thus, certain leaders have promoted an aggressive discourse against journalists and the media, initially being disseminated by influencers and trolls on social networks and finally reaching even ordinary users [32].

In addition, in the case of journalists, women and minorities are the main targets of online harassment, including death threats, the disclosure of personal data, or gender-based violence [33].

Sometimes, the harassment of journalists using social networks takes on the very characteristics of hate speech [34], i.e., attacking a person or group on the basis of the features of their identity [35]. Although this concept has a variety of meanings [36], such speech is usually understood to be messages that incite violence and hate against a group [37] and that communicate an ideology, usually via stereotypes according to specific features of a gender, religion, race, or disability [38]. Therefore, hate speech is best understood as a subset of broader bullying and harassment behaviors, and it is characterized by explicit references to identity-based traits.

Indeed, many women journalists report sexist and misogynistic attacks on social networks [39], which target them more frequently than their male colleagues [40]. Such harassment in the digital sphere, particularly against those covering political and gender-related topics [41,42], often mirrors the abuse that they endure in other contexts [43,44], thereby reinforcing the structural inequalities that persist in journalism and society as a whole [45].

Thus, hate speech directed at women journalists can be considered to be a specific type of harassment that includes sexist or misogynistic comments via which they are criticized, attacked, marginalized, stereotyped, or threatened on the basis of their gender or sexuality. This is different from the ordinary harassment of journalists seen on social networks [46,47].

The digital harassment of women journalists has thus become one of the main objects of interest in this field, although the limited availability of empirical research hinders the development of effective responses [48].

Most analyses have been carried out through surveys [40,49], focus groups [50], or semi-structured in-depth interviews with professionals in different countries [17,19,21,22,41,47,51–54]. Studies have also been carried out in a number of specialized areas of journalism—for example, sports [35,43] or technology [15]. However, quantitative studies based on social media analysis, particularly in political contexts and in Spain, remain scarce [55,56].

In short, the relationship between social media and journalism reveals a fundamental paradox: while media organizations encourage their staff to maintain an active presence on these platforms [57], social media often functions as an echo chamber that heightens the risk of harassment. This dual nature of digital tools means that, although they offer increased visibility and opportunities for audience engagement, they also expose journalists to harmful interactions [45,58-60].

In this context, this study analyzes the occurrence of harassment and hate speech directed at journalists and the media against the backdrop of a process with a significant news impact, namely the Spanish general elections of July 23 (23-J). On the basis of this general objective, the following research questions (RQs) are posed.

RQ1. What is the intensity of the harassment and hate speech directed at Spanish journalists on social networks?
RQ2. What types of harassment are the most common, and what topics are they related to?
RQ3. Is gender a factor in this type of online harassment? Does this gender-related factor have a political orientation?

## 2. Materials and Methods

Twitter (currently called X) was selected for this study because it is the social network that journalists use most in their daily work, far more than Facebook or Instagram [5]. This social network has also been the subject of much research in the communication field because of its special relationship with journalism [11,61]. In addition, Twitter is one of the favorite means adopted by politicians to disseminate their proposals and campaign messages during electoral processes [62]. At the time of writing, the platform has over four million accounts in Spain; one in five of these is considered active, with at least one message posted during the preceding 2 months [63]. Users who are men slightly outnumber those who are women on this network, at 32% versus 28%; however, the majority (40%) do not specify their gender.

In total, 200 profiles with a high level of activity and importance were analyzed; half of these belonged to media outlets and half to journalists. The media outlet accounts were drawn from the Iberian Digital Media Map catalog [64], which contains 4039 records, among which the 100 with the greatest number of followers were selected. For journalists, the lists of active professionals available on this social network were analyzed, and, from these, a preliminary list of 568 users was drawn up. Of these, the 100 accounts with the greatest number of followers were ultimately selected.

To quantify harassment and hate messages on social networks, responses to messages posted by the selected journalists and media outlets in the days immediately preceding and following the 23-J elections—specifically between 18 July and 31 July (14 days)—were analyzed. In total, 60,684 user responses to messages originally written by journalists (30,895) and media outlets (29,789) were analyzed.

A two-step approach was adopted to identify harassment and hate speech. First, using the X/Twitter Academic API, the responses directed at these users were collected, limited to a maximum of 100 responses per user, in accordance with API restrictions in effect at the time of data collection prior to its discontinuation. Second, once the responses were collected, they were translated into English, and the machine learning model (large language model [LLM]) cardiffnlp/twitter-roberta-base-hate-latest from the Hugging Face machine learning ecosystem was applied to identify and categorize instances of hate speech. In addition to these automatic classifications, categories for the users' genders and specific types of harassment or hate speech were incorporated manually. Although there are hate speech detectors for Spanish [65,66], a more robust and current English-language model was selected.

Thus, the texts included in the sample were first analyzed automatically and then reviewed manually to refine the selection and correct errors and false positives, messages whose posters had not been confirmed, and adversarial content not strictly constituting harassment or hate.

Note that although the quantity of toxic words may indicate toxicity, this does not necessarily mean that the intention of the tweet is to attack, and other contextual aspects should be considered [39].

The Doccano software (v. 1.6) was used for data labeling, and an exploratory data analysis was performed using the Python programming language (v. 3.11.6). In this way, a final sample of 1478 responses with harassment or hate content (2.42%) was established.

Building upon the previous studies available, an inductive method based on an exploratory analysis of 10% of the items in the sample was used to categorize the messages. On the basis of this analysis, seven categories to measure harassment and hate speech directed at journalists were ultimately established: (1) insults, (2) political hate, (3) hate directed at media outlets, (4) racial discrimination and hate, (5) sexism and misogyny, (6) homophobia/LGBTQ+ phobia, and (7) other. The categories used in the analysis were not mutually exclusive, as individual messages may contain multiple forms of aggression (e.g., a sexist comment that is also politically motivated). Rather than imposing a single-label classification, messages were coded according to all relevant categories that they displayed, including both general insults or political hatred and specific forms of hate speech targeting identity-based characteristics (e.g., sexism, racism, homophobia, etc.). Likewise, the political orientation of these responses was identified on the basis of a semantic analysis of the qualifiers in the text (e.g., "fascist" ["facha"], "red" ["rojo"], etc.). We also analyzed whether they were completely or partially directed at journalists and media outlets or whether, on the contrary, the responses used such messages to attack third parties. All coding was carried out by a single researcher using an intra-rater reliability system, achieving an accuracy level above 90%.

For the analysis of the results, we first identified outliers, defined as samples in the dataset that differed exceptionally from the majority of the data [67]. In this case study, the standard deviation was not calculated because the dataset analyzed did not follow a normal distribution. Therefore, the chosen measure was the interquartile range (IQR), with 1.5 times the IQR being considered an extreme value.

The Kruskal–Wallis test was employed, as the data did not meet the conditions for parametric analysis (i.e., normality), although they did show equal variance. Effect sizes for differences between user categories were measured using the η² test, which allowed for the evaluation of the observed similarities and differences among the groups.

## 3. Results

*3.1. Political Hate or Insults*

The analysis of user responses to messages that journalists and media outlets posted on Twitter in the context of the 23-J election campaign indicated that 2.42% of these messages ($n$ = 1478) included content identifiable as harassment or hate speech. The targets of these attacks were varied but always had a strong connection to politics. Of the messages including harassment, 27.13% ($n$ = 401) were specifically directed at journalists and media outlets, while 34.17% ($n$ = 505) attacked them along with other targets, often in the context of the elections. Conversely, 34.71% ($n$ = 513) of hostile responses attacked third parties rather than journalists or media outlets. In total, only one in three of the accounts analyzed did not receive any responses including hate content during the 2-week sample period.

Focusing on the messages aimed directly or partially at journalists or media outlets, nearly half (46.69%, $n$ = 494) expressed clear political hate, which aligned with the nature of the electoral context. Qualifiers such as "fascist" ["fascista"], "Nazi," "communist" ["comunista"], or "terrorist" ["terrorista"] were the most prominent in an atmosphere of strong bipartisanship and high polarization. The second most common category was insults and generic slurs (32.61%, $n$ = 345).

By contrast, there was much less content that specifically constituted hate speech, i.e., attacks based on membership to a specific group or on the attribution of its identity (Table 1). In total, elements constituting discrimination or racial hate were identified in 7.18% of the messages analyzed ($n$ = 76), whereas 5.77% ($n$ = 61) included sexist or misogynistic qualifiers. In addition, 3.31% ($n$ = 35) of the messages could be categorized as homophobic/LGBTQ+-phobic. Finally, hate specifically targeting the media and journalists—with qualifiers such as "liars" ["mentirosos"], "fake news" ["bulos"], or "corrupt" ["corruptos"]—appeared in 1.98% of the messages ($n$ = 21).

Regarding the account types, individual journalists' profiles received twice as many hostile messages compared with the institutional accounts of the media outlets where they worked. The profiles that received the greatest number of messages including harassment belonged, almost entirely, to media directors and newspaper opinion leaders or to press columnists, presenters, or talk-show hosts on political programs (Table 2). Focusing on outliers, i.e., those cases that exceeded at least 1.5 times the interquartile range, the accounts including the most harassment were those of @AlfonsoRojoPD, @javiernegre10, @pacomarhuenda, @AntonioMaestre, @estherpalomera, and @AntonioRNaranjo.

Analyzing the political orientation of the messages also revealed significant data. Of the 935 messages including harassment and hate received by journalists and media outlets, 44.04% ($n$ = 425) came from left-wing positions, compared with 32.02% ($n$ = 309) from right-wing positions. Messages whose political stance was unclear or could not be determined accounted for 23.94% ($n$ = 231). To determine the political orientation of a message, the political nature and the slant of the slurs contained in the response were analyzed. Therefore, considering the data on the political stances of harassment or hate messages and their addressees, the number of messages from the left that were directed at journalists working in media outlets affiliated with the right was somewhat higher than in the opposite case.

Table 1. Harassment of journalists and the media on Twitter.

| | Total Journalists | | Men | | Women | | Media Outlets | |
|---|---|---|---|---|---|---|---|---|
| | Total | (%) | Total | (%) | Total | (%) | Total | (%) |
| Insult | 251 | 36.01% | 183 | 37.81% | 68 | 31.92% | 94 | 26.48% |
| Political hate | 346 | 49.64% | 244 | 50.41% | 102 | 47.89% | 148 | 41.69% |
| Hate directed at the media | 13 | 1.87% | 11 | 2.27% | 2 | 0.94% | 6 | 1.69% |
| Racial discrimination/hate | 15 | 2.15% | 6 | 1.24% | 9 | 4.23% | 58 | 16.34% |
| Sexism/misogyny | 41 | 5.88% | 13 | 2.69% | 28 | 13.15% | 20 | 5.63% |
| Homophobia/LGBTQ+phobia | 22 | 3.16% | 19 | 3.93% | 3 | 1.41% | 13 | 3.66% |
| Other | 9 | 1.29% | 8 | 1.66% | 1 | 0.47% | 16 | 4.5% |
| TOTAL | 697 | 100% | 484 | 100% | 213 | 100% | 347 | 100% |

Source: Authors' own creation.

Table 2. Journalists who received the most harassment and hate (number of responses).

| | Journalist | Gender | Number | | Journalist | Gender | Number |
|---|---|---|---|---|---|---|---|
| 1 | AlfonsoRojoPD | M | 53 | 11 | JorgeBustos1 | M | 18 |
| 2 | javiernegre10 | M | 48 | 12 | RosaVillacastin | F | 18 |
| 3 | pacomarhuenda | M | 40 | 13 | sandrasabates11 | F | 17 |
| 4 | AntonioMaestre | M | 40 | 14 | LuciaMendezEM | F | 16 |
| 5 | estherpalomera | F | 36 | 15 | EstefaniaJoaq | M | 15 |
| 6 | AntonioRNaranjo | M | 28 | 16 | crpandemonium | M | 15 |
| 7 | ErnestoEkaizer | M | 25 | 17 | _InakiLopez_ | M | 15 |
| 8 | pedroj_ramirez | M | 22 | 18 | JesusCintora | M | 14 |
| 9 | carloscuestaEM | M | 20 | 19 | susannagriso | F | 13 |
| 10 | MariaJamardoC | F | 20 | 20 | PilarGGranja | F | 13 |

Source: Authors' own creation.

*3.2. The Gender Dimension*

In addition to a clear political dimension, the data analyzed also indicated that there was a gender dimension when it came to harassment and hate speech directed at women journalists—not generally, but in some specific issues identified in the details of the data analyzed.

On the one hand, the analysis of the overall data and of all the responses analyzed did not show statistically significant differences in the amount of harassment and hate received by women journalists compared with that received by men (Figure 1). The Kruskal–Wallis test (0.0413) indicated no difference in the median percentage of hate messages between the two groups, and, given that the *p*-value was high (0.8389) and the effect size was small ($6.49 \times 10^{-5}$), the statistical difference, if any, would not have been significant either.

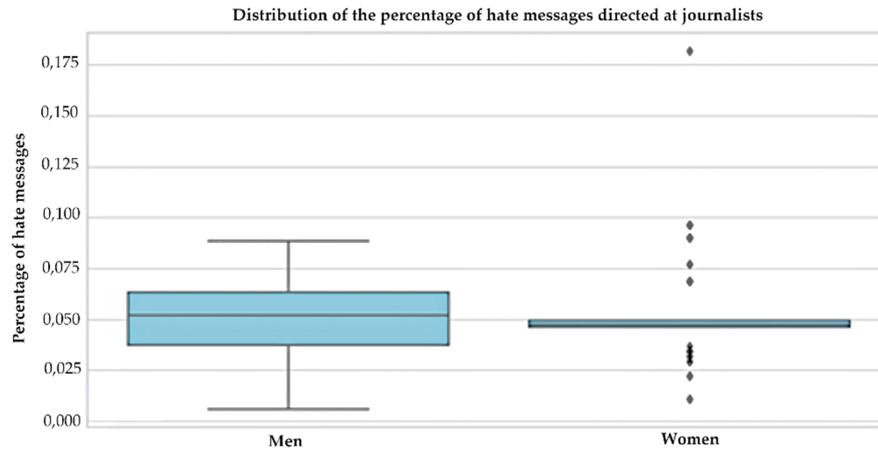

**Figure 1.** Hate-containing responses to journalists (by gender). Source: Authors' own creation.

On the other hand, messages that included insults and political hate were distributed with little variation between men and women and were influenced more by the political orientation ascribed to the recipient rather than their gender. As seen in the scatterplots (Figure 2), there was not a greater number of outliers for women when all the messages analyzed were considered. Although women journalists did appear in the scatterplot outside of the regression result, they received too little content containing harassment or hate to be seen as outliers.

However, when we excluded the two general categories (insults and political hate) and focused specifically on the categories that could be qualified as hate speech, the data suggested that the proportion of sexist messages received by women (13.15%, *n* = 28) was much higher than that received by men (5.88%, *n* = 41). Because of the small total number of responses identified in this subcategory compared with the sample as a whole, the results could not be considered sufficiently significant; however, they do indicate a clear trend.

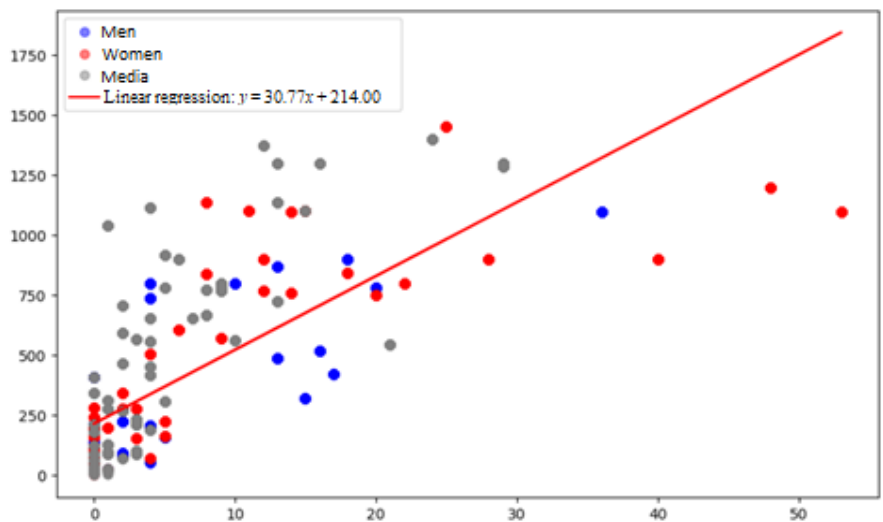

**Figure 2.** Harassment and hate directed at media outlets and journalists (by gender). Source: Authors' own creation.

Moreover, if we look at the political orientation of the harassment and hate received, a clear gender dimension also emerges. In this regard, the overall number of harassment messages directed at journalists and the media was somewhat higher for positions identified as left-wing. Despite this, the Kruskal–Wallis test (Figure 3) did not indicate significant differences in the medians between the groups when it came to these types of messages either.

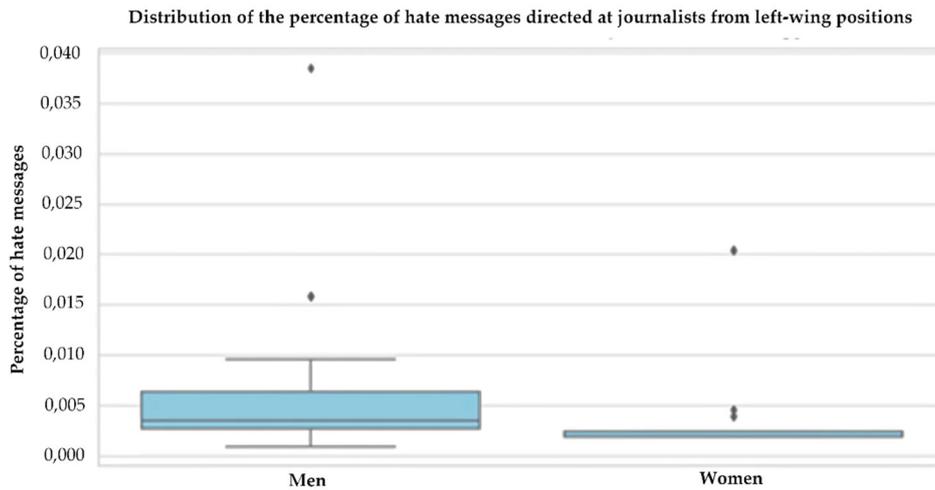

**Figure 3.** Hate-containing responses to journalists from left-wing positions (by gender). Source: Authors' own creation.

In contrast, the gender dimension was clear and highly significant in messages written from a right-wing stance (Figure 4). For these, the Kruskal–Wallis statistic was 45.7373, and the associated *p*-value was $1.3522 \times 10^{-11}$, which, together with a moderate effect size (0.1675), indicated a significant difference between men and women in terms of the median percentage of hate messages.

In other words, in a context in which journalists and media outlets receive slightly more harassment and hate on X/Twitter from left-wing positions, women journalists receive this harassment to a particularly significant degree from individuals with right-wing positions, which supports the notion of an ideological dimension of gendered digital harassment.

Therefore, in terms of gender, although the total number of sexist messages and hate messages was lower than the proportion of political hate and insults, it is noteworthy that sexist messages were received to a much greater extent by women and were sent significantly more often by users with extremist or right-wing populist positions.

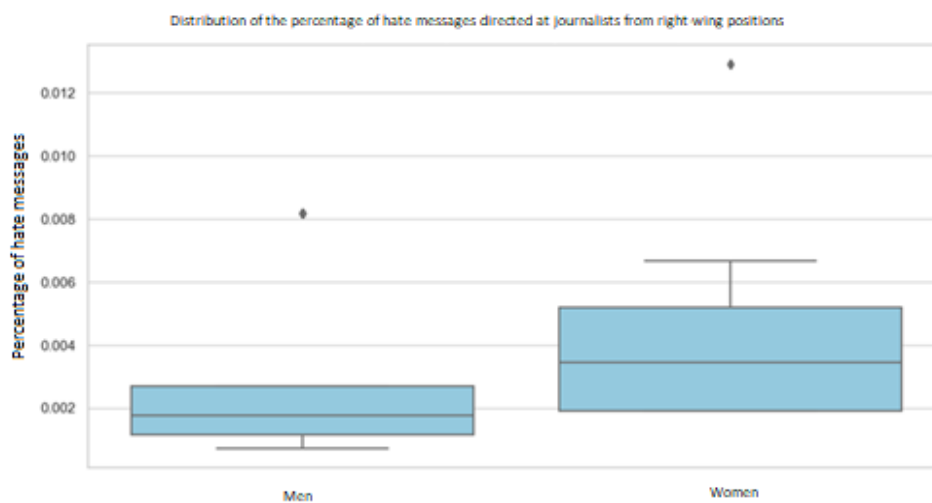

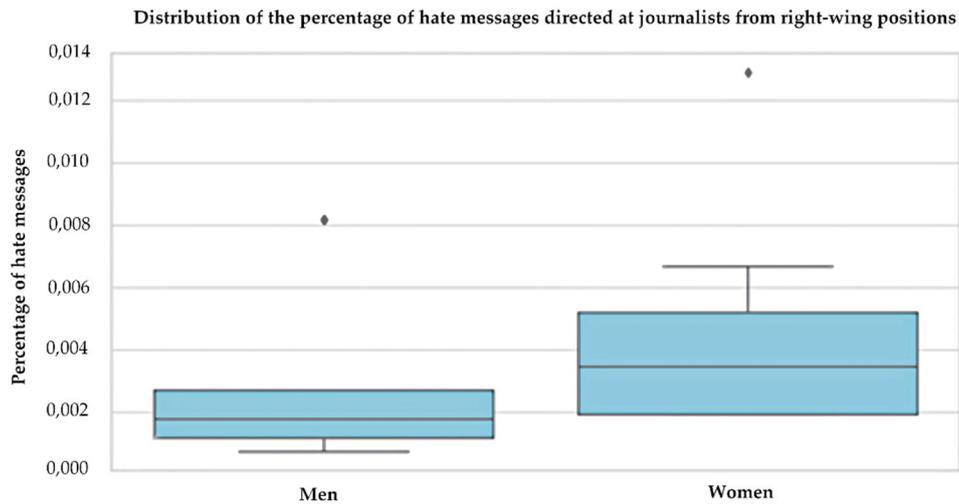

**Figure 4.** Hate-containing responses to journalists from those with right-wing positions (by gender).

## 4. Discussion

This analysis of harassment and hate in the responses received to messages that journalists and the media posted in the context of the 23-J general elections shows, primarily, the prevalence of political hate directed at journalists. The data indicate that, during the electoral period, there was a "shoot the messenger" effect, where, in a climate of strong polarization, journalists were the targets of attacks from one political extreme or the other depending on their support—real or ascribed—of one of the dominant political stances. While online harassment may contribute to broader patterns of silencing and self-censorship, this study finds that such abuse is largely ideologically driven, with a highly charged political environment emerging as a central factor in the hostility that journalists face—fueling attacks and reinforcing antagonistic dynamics in public discourse. These results are consistent with those obtained by other previous studies [17,41], which have even identified organized harassment and persecution campaigns against journalists, sometimes spilling out beyond the digital sphere [50], and they underline the relevance of the political context when it comes to the intensity and nature of harassment.

Aside from political hate, with regard to messages specifically identifiable as hate speech—that is, those referring to membership to a specific group—sexism or gender-based digital harassment was the most common type of harassment. The data suggest that this type of harassment of women is common, although the sample analyzed, extracted at a time of strong political mobilization, provided insufficient evidence to establish statistically significant differences. A possible future line of research could be to try to corroborate what these data suggest using a broader temporal sample obtained at times of less political polarization, which would allow this trend to emerge.

Likewise, a significant statistical difference was found when it came to digital harassment and hate speech from those in right-wing positions directed at women. Thus, in a climate in which the overall orientation leaned in the opposite direction, the number of hostile remarks directed at women journalists that were found in right-wing users' responses was significantly higher. Thus, although, at an aggregate level, the gender-based harassment of women journalists was not found to be more common, it was found that more harassment with a gender dimension came from users with certain political positions, which is in line with public attacks made by some populist leaders [28,29]. The results obtained also emphasize the personal dimension of digital harassment, as journalists' individual accounts received more insults and hate than institutional media accounts. This point calls for reflection regarding journalists' increased exposure on social networks and the failure of media outlets' codes of conduct for their use. Given that the individual accounts used by the vast majority of journalists to share messages exist somewhere in the grey area between their personal and professional spheres, the intensity of attacks on individuals, far surpassing that experienced by institutions, also indicates the intention to influence their activity and silence them [30]. This targeted hostility toward individual journalists contributes to a hostile environment that can ultimately fuel self-restraint and discourage open participation in public debate.

Finally, although the overall impact was limited, hate speech and digital harassment directed at journalists, at least in their less extreme forms, can be considered to be systemic across social networks [20,25], particularly when it comes to highly polarized contexts such as electoral processes. It should also be noted that the study sample was selected to include the 100 journalists with the greatest number of followers; as opinion leaders with a great deal of public exposure, they are at greater risk of being harassed on social networks [68]. In this context, it can be said that social networks should be understood not only as interactive spaces but also as settings that can amplify hostility and hate speech, negatively influencing journalistic practices and democratic quality.

Therefore, although online harassment has not usually been considered important [40], or has even been normalized as part of the price to pay to have a greater audience impact [51], the media must continue to develop preventive and palliative mechanisms to deal with it [21], providing effective legal support and organizational measures to address the psychosocial consequences of harassment [31]. Moreover, the need for protective tools and policies, the development of which has varied from country to country [33,52], must extend beyond the level of the corporate media accounts to which they are usually limited and cover the individual accounts that most journalists use to share their messages in the public domain. Although there are legal developments, such as convictions for harassment and defamation, thus far, legislative gaps limit such protection, owing either to the existence of restrictive definitions of harassment, the absence of gender-sensitive policies, or a reliance on victim reporting [69].

Even though these professionals often operate within the grey area between personal opinion and professional activity, their exposure to the public, as well as much of their work, stems from working for the media, so the individual harassment that they experience is inextricably linked to the social dimension of their work.

This study also has some limitations. On the one hand, the analysis focused on the 23-J general election period, which could have biased the results toward a highly polarized context. This limits the generalizability of the conclusions to less politicized periods. Moreover, the sample included only the 100 journalists with the greatest number of followers, which may not reflect the experiences of harassment and hate toward journalists with less exposure. Finally, the content of the original message that triggered the response was not analyzed. Regardless of the validity of the analysis of the responses in and of themselves, the original messages' characteristics—including their truthfulness, tone, and political leanings—may have had an effect on the types of interactions that they produced. Future lines of research may include a longitudinal analysis of hate speech over an extended period of time, a sample encompassing all Spanish media professionals, and a comparison of the occurrence of harassment on personal versus professional accounts.

**Funding:** This work is part of the research project TED2021-130810B-C22, funded by the State Research Agency (AEI) of the Ministry of Science and Innovation (MICIN) of Spain 10.13039/501100011033 and by the European Commission Next Generation EU/PRTR. It is also part of the scientific output of the Consolidated Research Group of the Basque University System, Gureiker (IT1496-22).